\documentstyle[osa,aps,prl,epsfig]{revtex}
\begin{document}
\twocolumn[\hsize\textwidth\columnwidth\hsize\csname
@twocolumnfalse\endcsname


\title{Observation of First-Order Metal-Insulator Transition without Structural Phase Transition in VO$_2$}
\author{Yong-Sik Lim,$^1$ Hyun-Tak Kim,$^{2\ast}$ B. G. Chae,$^2$ D. H. Youn,$^2$ K. O. Kim,$^2$ K. Y. Kang,$^2$ S. J. Lee,$^3$ K. Kim$^3$}
\address{$^1$Department of Applied Physics, Konkuk University, Chungju 380-701,
Korea\\$^2$Telecom. Basic Research Lab, ETRI, Daejeon 305-350,
Korea\\$^3$School of Chemistry, Seoul National University, Seoul
151-742, Korea}
\date{January 26, 2003}
\maketitle

\begin{abstract}
An abrupt first-order metal-insulator transition (MIT) without
structural phase transition is first observed by current-voltage
measurements and micro-Raman scattering experiments, when a DC
electric field is applied to a Mott insulator VO$_2$ based
two-terminal device. An abrupt current jump is measured at a
critical electric field. The Raman-shift frequency and the
bandwidth of the most predominant Raman-active A$_g$ mode, excited
by the electric field, do not change through the abrupt MIT,
while, they, excited by temperature, pronouncedly soften and damp
(structural MIT), respectively. This structural MIT is found to occur secondarily.\\
PACS numbers {71.27.+a, 71.30.+h, 78.30.Hv}\\ \\
\end{abstract}
]


Since Mott first predicted an abrupt first-order metal-insulator
transition (MIT) without structural phase transition driven by a
strongly correlated electronic Coulomb energy \cite{Mott}, the MIT
has not been found in experiments
\cite{Imada,Fazekas,Morin,McWhan,Okazaki,Limelette}. Observations
of first-order MITs for a Mott insulator \cite{Rice} have been
achieved by temperature and pressure, and are always accompanied
with a structural phase transition from the monoclinic structure
of an insulator to the tetragonal structure of a metal
\cite{Morin,McWhan,Okazaki}. This complicates the basic mechanism
of the Mott transition and divides it into two major mechanisms:
the electron-phonon interaction and the electron-electron
correlation. Moreover, whether a MIT near a Mott insulator is
continuous or abrupt has long been a problem
\cite{Imada,Fazekas,Rice,Wentzcovitch}.

We have observed a MIT with an abrupt jump of driving current,
when an electric field is applied to two-terminal devices with an
interval of 5 $\mu$m fabricated on an epitaxial VO$_2$ film
\cite{HTKim}. It is thus imperative to investigate whether the
abrupt MIT is accompanied by a structural phase transition. To
observe a structural change, a micro-Raman scattering experiment
rather than neutron diffraction is suitable to obtain experimental
data as the size of the neutron beam cannot be reduced within 5
$\mu$m.

In this letter, we report an abrupt first-order MIT, which does
not undergo a structural phase transition, observed by
current-voltage measurements and micro-Raman scattering
experiments with application of a DC electric field to a
two-terminal device fabricated on a Mott insulator VO$_2$ film.

Vanadium dioxide, VO$_2$, has been studied extensively because it
displays an abrupt MIT at a critical temperature $T_c\approx$340 K
and a slight lattice distortion from a monoclinic structure with
space group C$^{5}_{2h}$ for the insulator phase to a tetragonal
rutile structure with space group D$^{14}_{4h}$ for the metal
phase. From the group theory for VO$_2$, five modes (A$_{1g}$,
B$_{1g}$, B$_{2g}$, E$_g$, A$_{2g}$ (silent mode)) to the high
temperature phase and eighteen modes (nine A$_g$ and nine B$_g$)
to the low temperature phase are revealed as Raman active modes
\cite{Srivastava}. Schilbe found all permissible Raman modes of
the insulator phase, except one, and also estimated four possible
Raman-mode peaks with broad bandwidth of the metal phase above
$T_c$ \cite{Schilbe}.

\begin{figure}
\vspace{-0.0cm}
\centerline{\epsfysize=9.5cm\epsfxsize=8cm\epsfbox{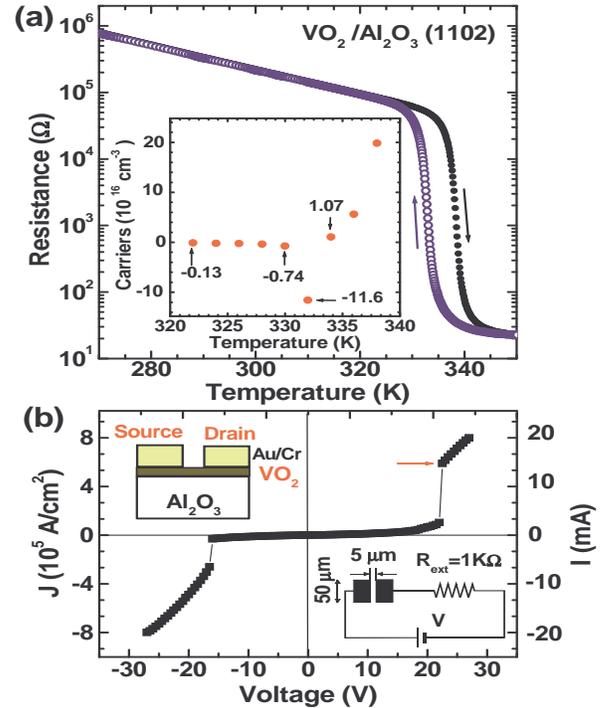}}
\vspace{0.2cm} \caption{(a), Temperature dependence of resistance
of VO$_2$ film I. An abrupt MIT at a critical temperature
$T_c{\approx}$340 K is clearly shown. The inset shows the number
of carriers obtained by Hall measurements. A change of carriers
from holes to electrons is shown at 332 K. The minus sign
indicates that the carriers are holes. (b) Current-density J vs.
voltage V curve measured by VO$_2$ based two-terminal device I. An
abrupt MIT is shown. The inset is a circuit with 1 K$\Omega$ and a
layout of a two-terminal device.}
\end{figure}

\begin{figure}
\vspace{-0.5cm}
\centerline{\epsfysize=9.5cm\epsfxsize=9.0cm\epsfbox{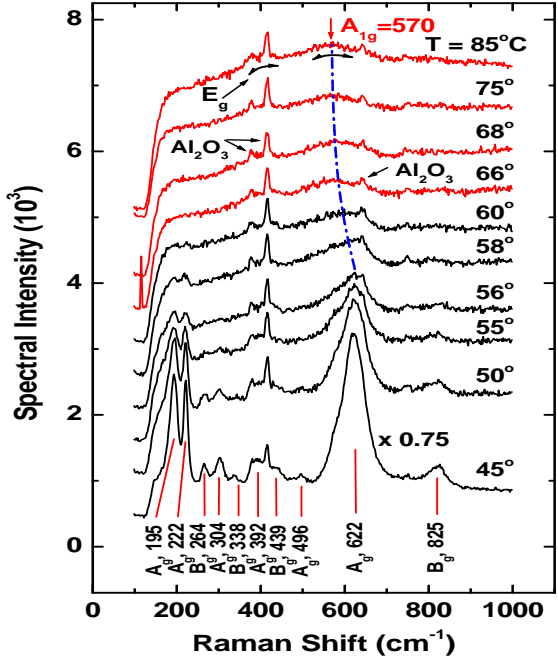}}
\vspace{0.0cm} \caption{Temperature dependence of Raman spectra
measured by VO$_2$ based two-terminal device II. Above T =
66$^{\circ}$C, an Raman active A$_{1g}$ mode near 570 cm$^{-1}$ of
metal phase appears (blue and blue-dot line). The A$_{1g}$ mode is
heavily damped (red spectra). Below T = 45$^{\circ}$C, Raman
active modes of insulator phase are resolved by peak position.}
\end{figure}

Epitaxial thin films of VO$_2$ have been deposited on (1102)
Al$_2$O$_3$ substrates by laser ablation \cite{Chae,Youn}. The
thickness of the VO$_2$ films was about 900$\AA$. For two-terminal
devices, Ohmic-contacted Au/Cr electrodes on VO$_2$ films with a
width of 50 $\mu$m and a length of 5 $\mu$m (inset of Fig. 1b)
were patterned by photo-lithography and lift-off. All devices used
in this research had the same dimensions. Raman spectra were
measured by exposing an Ar laser beam to VO$_2$ between
electrodes. A 17 mW Ar ion laser with a 514.5nm line in a
micro-Raman system (Renishaw 2000) with a spectral resolution less
than 2 cm$^{-1}$ was employed. The Raman system was also equipped
with an integral microscope (Olympus BH2-UMA). When the Raman
spectra are measured, the current after the abrupt MIT was limited
to compliance current to prevent the device from possible damage
due to excess current. Even though the compliance current was
extended from 2 to 100 mA (measurable maximum current of our
system), an abrupt current jump was always observed at a MIT
voltage V$_{MIT}$  = 10 $\sim$ 11V. Current I vs. voltage V curves
were measured by a precision semiconductor parameter analyzer
(HP4156B).

Figure 1a shows the temperature dependence of resistance of VO$_2$
film I. An abrupt MIT and hysteresis are shown near
$T_c\approx$340 K (68$^{\circ}$C). This is consistent with
previous measurements \cite{Natale,Borek}. It was proposed that
this abrupt MIT arises from the structural phase transition from
monoclinic below $T_c$ to tetragonal above $T_c$
\cite{Morin,Okazaki}. Hall measurements reveal that the number of
hole carriers increases with increasing temperature by
$T_c\approx$340 K, and that electron and hole carriers coexist
near $T_c$ (inset of Fig. 1a). Owing to mixing of an electron Hall
voltage and a hole Hall voltage, the number of carriers at
temperatures from 332 to 340 K cannot be exactly determined. The
number of hole carriers at $T_c$ is expected to be
n$_c\approx$3$\times$10$^{18}~$cm$^{-3}$ (0.018$\%$ of $d$-band
charges) from the Mott criterion causing the first-order MIT
\cite{Mott}, based on an exponential decrease of resistance
(increase of carrier) with increasing temperature. In the metal
regime above 340 K, the major carriers are electrons (inset of
Fig. 1a). Fig. 1b shows a current density J vs. voltage V curve
measured by two-terminal device I fabricated on VO$_2$ film II. An
abrupt jump at 21.5V and Ohmic behavior (metal characteristic)
above 21.5V are exhibited. The current density after the current
jump is about $J\approx$~6$\times$10$^5$ A/cm$^2$ measured in a
circuit with 1 K$\Omega$, which corresponds to a value obtained in
a dirty metal. These are typical characteristics of a first-order
MIT.

\begin{figure}
\vspace{0.0cm}
\centerline{\epsfysize=10.0cm\epsfxsize=8.0cm\epsfbox{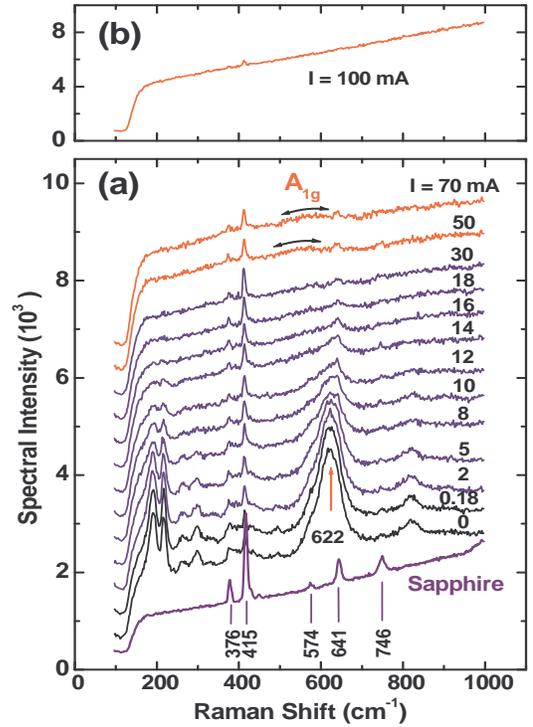}}
\vspace{0.0cm} \caption{(a) Compliance-current dependence of Raman
spectra measured by VO$_2$ based two-terminal device II. The MIT
transition voltage V$_{MIT}$ = 10 $\sim$ 11 V at I = 30 mA (blue
spectra), the applied voltage V = 15 V at I = 50 mA, and V = 19 V
at I = 70 mA, (b) V = 25 V at I = 100 mA between the
two-terminals. Above I = 16 mA, A$_g$ modes of the monoclinic
insulator phase are not shown. The Raman lines of Al$_2$O$_3$
substrates are also shown on the bottom of the figure. The spectra
were measured at room temperature.}
\end{figure}

Figure 2 shows Raman spectra measured at temperatures between 45
and 85$^{\circ}$C during a cooling cycle with VO$_2$ based
two-terminal device II. The spectra are superimposed on the
contribution of the Al$_2$O$_3$ substrate with sharp peaks near
376, 415, 574, 641, and 746 cm$^{-1}$ (bottom of Fig. 3a). The
polarized directions of incoming and outgoing light propagating to
the direction of [1102] are all the same. The Raman spectra
display: (i) features of a distorted monoclinic rutile structure
with narrow line shapes below 50$^{\circ}$C, (ii) a typical
characteristic in the coexistence of semiconductor and metal
phases in the range of 50$\sim$60$^{\circ}$C, and (iii) a broad
peak around 570 cm$^{-1}$ as a Raman active mode of metal phase of
VO$_2$ above $T$ = 60$^{\circ}$C (dash-and-dotted line in Fig. 2).
Although there have been some reports where Raman active modes are
not observed in a VO$_2$ film above $T_c$ \cite{Foster,Petrov},
our result is consistent with that of previous investigations in a
bulk VO$_2$ \cite{Srivastava,Vikhnin}. Compared with the previous
measurement \cite{Schilbe} of VO$_2$ at 83 K, we can resolve
almost all allowable Raman active modes (6 A$_g$ and 4 B$_g$) at
room temperature, except for a very weak A$_g$ mode near 595
cm$^{-1}$ and a strong A$_g$ mode near 149 cm$^{-1}$, which is
suppressed by a cutoff filter in our measurement system. Both peak
position and assignment of the symmetry modes are denoted at the
bottom of Fig. 2. All B$_g$ modes were weakly measured in our
polarization configuration to satisfy the Raman selection rule. As
temperature increases, the intensities of narrow and strong A$_g$
mode peaks decrease and bandwidths seem to broaden, while a new
broad Raman peak near 570 cm$^{-1}$ appears. Above $T$ =
66$^{\circ}$C, Raman active modes of the monoclinic phase are not
shown; instead, two broad Raman peaks superimposed on the sharp
Al$_2$O$_3$ modes, strong near 570 cm$^{-1}$ and weak near 400
cm$^{-1}$, appear and are clearer with increasing temperature (top
of Fig. 2). These broad peaks were assigned to phonon modes with
A$_{1g}$ and E$_g$ symmetries, respectively
\cite{Srivastava,Schilbe}. In particular, the broad A$_{1g}$-mode
peak near 577 cm$^{-1}$  becomes obvious with increasing
temperature from 58$^{\circ}$C. This may be attributed to
simultaneous changes of the crystal structure and the
conductivity.

Figure 3a shows the compliance-current dependence of Raman spectra
observed at the current jump of a MIT voltage V$_{MIT}$  = 10
$\sim$ 11V for VO$_2$ based two-terminal device II. With
increasing of the compliance current, A$_g$ and B$_g$ modes in the
spectra vanish without any change in their peak positions and
bandwidths, which indicates that the features of the monoclinic
phase do not change. From I = 16 to 30 mA (blue lines in Fig. 3a),
the peaks of Raman active mode of VO$_2$ are not found. The ratio
of background levels of Raman spectra near 1000 cm$^{-1}$ for the
metal phase at I  = 16 mA to the insulator phase at I = 0 mA is
about 1.23, which is quite close to the value of 1.20 obtained
from the temperature dependence by considering the Boltzmann
correction between 26$^{\circ}$C at room temperature and
68$^{\circ}$C near $T_c$ in Fig. 2. This indicates that the
measured phase has already become metal. Moreover, the ratio at I
= 100 mA (Fig. 3b) is as high as 3.11, which is due to the
increase of the number of conduction carriers excited by the
electric field. In particular, the spectra of I = 50 and 70 mA
measured at 15V and 19V in the Ohmic regime (over V$_{MIT}$  = 10
$\sim$ 11V), respectively, show a broad peak near 570 cm$^{-1}$
(arrow in Fig. 3a). The broad peak is a secondary effect produced
by Joule heating due to high conduction after MIT, as described in
a following section. Note that the residual resistance is about
250 $\Omega$ in the Ohmic regime and the current density is about
$J \approx$ 1.1 $\times$ 10$^7$ A/cm$^2$ at I = 50 mA. Moreover,
the spectrum of 100 mA was also measured at V = 25V.

\begin{figure}
\vspace{-0.1cm}
\centerline{\epsfysize=9.3cm\epsfxsize=8cm\epsfbox{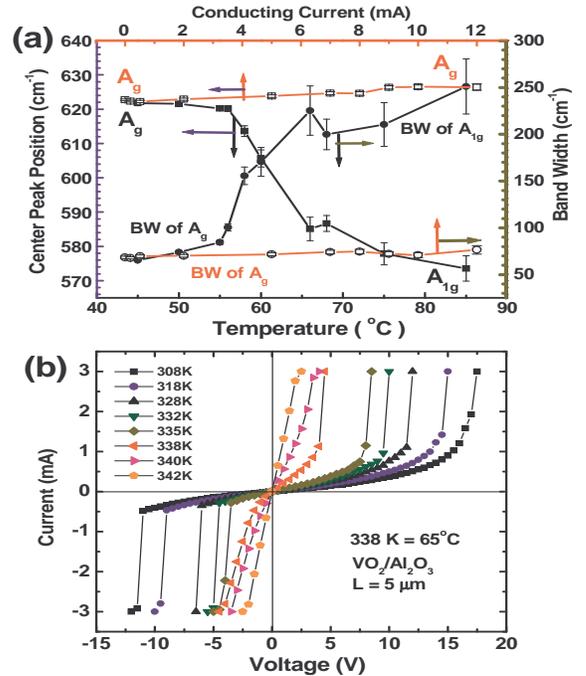}}
\vspace{0.0cm} \caption{(a) Temperature (filled symbol and black
line) and compliance-current (open symbol and red line) dependence
of a Raman A$_g$ mode near 622 cm$^{-1}$. A single Lorentzian
shape, taking into account the linear background contribution of
Al$_2$O$_3$ substrates, was assumed to fit the peak position
(square and black line) and the bandwidth (circle and red line) of
the Raman mode. BW means the bandwidth. (b) Temperature dependence
of the abrupt MIT observed at VO$_2$ based two-terminal device III
in a cryostat above room temperature. Near $T_c$, the abrupt MIT
disappears and Ohmic behavior appears. The MIT voltage decreases
with increasing temperature.}
\end{figure}

To compare the abrupt MIT excited by temperature with that by an
electric field in detail, we investigate changes of the peak
position and bandwidth of the predominant A$_g$ mode near 622
cm$^{-1}$. When the A$_g$ mode is assumed to be only a single
Lorentzian shape, the shape spectra contain both the contributions
of the Al$_2$O$_3$ substrates, showing linear dependence (bottom
of Fig. 3a), and the new A$_{1g}$ mode near 570 cm$^{-1}$ (top of
Fig. 3a). Fig. 4a shows the temperature and the compliance-current
dependence of the position and bandwidth of the A$_g$-mode peak of
622 cm$^{-1}$. With increasing temperature, the peak position of
the A$_g$ mode downshifts and the A$_g$ mode changes to the
A$_{1g}$ mode (black filled square in Fig. 4a), and the initial
bandwidth of 65 cm$^{-1}$ broadens (black filled circle in Fig.
4a). Furthermore, for Lorentzian-fitting results of temperature
behaviors of A$_g$ modes near 195 cm$^{-1}$ and 222 cm$^{-1}$
below $T$ = 60$^{\circ}$C, it was found that the change of peak
positions ($<$ 2 cm$^{-1}$) and bandwidths ($<$ 3 cm$^{-1}$) was
slight, except for an increase from 11 cm$^{-1}$ to 20 cm$^{-1}$
of bandwidth of the A$_g$ mode near 222 cm$^{-1}$. Note that, in
the metal phase above $T_c$= 60 K, these A$_g$ modes should vanish
and feasible B$_{1g}$ mode could be active (B$_{1g}$ near 240
cm$^{-1}$ may be screened here) \cite{Schilbe}.

Meanwhile, for a MIT excited by an electric field, the position
and bandwidth of the A$_g$-mode peak near 622 cm$^{-1}$ is nearly
unchanged (open square and open circle with red lines in Fig. 4a,
respectively). At most, the bandwidth of the A$_g$ mode becomes
only 8 cm$^{-1}$ wide, and the peak position shifts by only 3
cm$^{-1}$ to the high frequency side. This indicates that the
abrupt MIT is basically independent of the structural phase
transition, although the MITs by temperature and pressure
accompany the structural phase transition
\cite{Morin,McWhan,Okazaki}.

Figure 4b shows the decrease of the transition voltage of the
abrupt MIT with increasing temperature and Ohmic behavior without
a current jump near $T_c \approx$ 340 K and over. If a structural
phase transition resulting from Joule heating occurred during the
Raman measurement, the abrupt current jump should disappear or the
transition voltage should also be remarkably decreased. Thus,
although the broad peak near 570 cm$^{-1}$ at I = 50 and 70 mA in
Fig. 3a is a sign of a structural phase transition by temperature,
the peak is a secondary effect because the spectra at I = 50 and
70 mA were also measured at V$_{MIT}$  = 10 $\sim$ 11V.

Since the electric field excitation can induce an abrupt MIT from
a Mott insulator by reducing only the repulsive Coulomb potential
between electrons as suggested by the Brinkman-Rice (BR) picture
\cite{Brinkman} and extended BR picture \cite{Kim}, it is thought
that carriers move like cold electrons on the Fermi surface
without lattice heating. This might cause a significant difference
in the field excitation from the thermal excitation and intense
photo-excitation, inevitably incurring a structural phase
transition via hot electron and/or hot phonon generation
\cite{Cavalleri,Becker,Rousse}.

In conclusion, by analysis of the behaviors of predominant
A$_g$-Raman-active modes, it was revealed that the abrupt
first-order MIT of VO$_2$ by a DC electric field occurs without a
structural phase transition, in contrast with that by temperature.
Thus, the MIT in strongly correlated materials occurs abruptly due
to a decrease of the on-site critical Coulomb interaction by an
excitation such as an electric field. Furthermore, the abrupt MIT
will provide very important clues for solving ongoing debates on
metallic characteristics near the MIT
\cite{Imada,Rice,Wentzcovitch} and for future device applications.

We would like to DS Kim for discussions. Konkuk University
supported YS LIM by a special grant for this research. YS Lim, SJ
Lee, K Kim measured Raman spectra. HT Kim (leader of this
research), BG Chae, DH Youn, KO Kim, KY Kang fabricated devices
and measured the temperature dependence of resistance, Hall
effect, I-V characteristics. In particular, YS Lim analyzed Raman
data in light of discussions with HT Kim.

\end{document}